\documentclass[12pt]{article}

\usepackage{amsmath}
\usepackage{amssymb}
\usepackage{graphicx}

\begin{document}

\title{\bf \mbox{Turbulence in nonabelian gauge theory}}

\author{
J{\"u}rgen Berges, Sebastian Scheffler, and D{\'e}nes Sexty\\[0.5cm]
Institute for Nuclear Physics\\
Darmstadt University of Technology\\
Schlossgartenstr. 9, 64289 Darmstadt, Germany}

\date{}
\begin{titlepage}
\maketitle
\def\thepage{}          

\begin{abstract}
Kolmogorov wave turbulence plays an important role for the thermalization process following plasma instabilities in nonabelian gauge theories. We show that classical-statistical simulations in SU(2) gauge theory indicate a Kolmogorov scaling exponent known from scalar models. In the range of validity of resummed perturbation theory this result is shown to agree with analytical estimates. We study the effect of classical-statistical versus quantum corrections and demonstrate that the latter lead to the absence of turbulence in the far ultraviolet.
\end{abstract}
 
\end{titlepage}

\renewcommand{\thepage}{\arabic{page}}

\section{Introduction}

Heavy-ion collision experiments explore strongly interacting matter starting from a transient far-from-equilibrium initial state. Available data from the Relativistic Heavy Ion Collider reveals remarkable, unexpected properties such as rapid apparent thermalization with robust collective phenomena. The explanation of these findings from quantum chromodynamics (QCD) provides a challenge for theory. There are theoretical indications that nonequilibrium dynamics related to plasma instabilities could be crucial for the understanding of the plasma's temporal evolution~\cite{Plasmainst,WongYangMills,Romatschke:2006nk,Berges:2007re}. These instabilities lead to exponential growth of occupation numbers in long wavelength modes on time scales much shorter than the asymptotic thermal equilibration time. 
After the fast initial period of exponential growth the dynamics slows down considerably. The system is still far from equilibrium at this stage and the subsequent slow evolution may be characterized by Kolmogorov wave turbulence~\cite{Zakharov}. 

In the turbulent range characteristic properties such as values for scaling exponents can be insensitive to the details of the underlying microscopic theory. The associated universality may lead to quantitative agreements for very different theories or energy scales. Wave turbulence has been studied in detail in weakly-coupled scalar theories in the context of early-universe reheating~\cite{Micha:2004bv,Berges:2008wm}. Though the underlying mechanisms are very different for QCD and for scalar inflaton dynamics, the subsequent evolution after an instability follows similar patterns: There a tachyonic or parametric resonance instability leads to the exponential growth of occupation numbers, followed again by a slow cascade~\cite{preheat1,preheat2,preheat3,preheat4,Berges:2002cz}. Wave turbulence in scalar theories is described by nonthermal distributions such as $n(p) \sim p^{-\kappa}$ with $\kappa = 3/2$ for cubic self-interactions in the presence of a field expectation value, or for quartic interactions $\kappa = 4/3$ ($5/3$) describing particle (energy) cascades~\cite{Micha:2004bv}. These distributions are distinct from a thermal high-temperature distribution $\sim p^{-1}$. The nonthermal power-law behavior can be obtained from stationary solutions of perturbative Boltzmann equations neglecting quantum corrections~\cite{Micha:2004bv}. A new class of nonperturbative scaling solutions, with strongly enhanced low-momentum fluctuations, has recently been found~\cite{Berges:2008wm}. A very detailed picture of the different scaling solutions emerges using nonequilibrium renormalization group techniques~\cite{Berges:2008sr}.  

In contrast to the detailed picture for scalar quantum theories, our knowledge about gauge theories is still in its infancies. Even at sufficiently high momenta, where perturbative estimates are expected to be applicable, conflicting statements can be found about the nature of power cascades following plasma instabilities~\cite{Arnold:2005ef,Mueller:2006up}. It has been noted that the value for the exponents characterizing scaling behavior for gauge theories could deviate from those known for scalar field dynamics. For instance, a detailed numerical study in Ref.~\cite{Arnold:2005ef} for SU(2) gauge theory in Coulomb gauge suggests the value two for the scaling exponent, which is not fully understood perturbatively and seems to have no corresponding value in scalar models. The estimate employs a separation of scales between suitably defined 'soft' and 'hard' momenta for sufficiently small characteristic running gauge coupling. The respective hard-loop effective theory of soft excitations is a nonabelian version of the linearized Vlasov equations of traditional plasma physics, which are based on collisionless kinetic theory for hard particles coupled to a soft classical field~\cite{Plasmainst}. Vlasov equations can also be derived starting from classical-statistical field theory~\cite{Berges:2004yj}. Therefore, they may also be considered as an approximation of the classical-statistical field theory limit of the respective quantum gauge theory~\cite{Romatschke:2006nk,Berges:2007re}. Classical-statistical lattice gauge theory provides a quantitative description in the presence of sufficiently large energy density or occupation numbers per mode.  

In this work we show that classical-statistical simulations following plasma instabilities in SU(2) gauge theory indicate a Kolmogorov scaling exponent known from scalar models (Sec.~\ref{sec:classicalsimulation}). More precisely, we demonstrate that the numerical simulation is consistent with $\kappa = 4/3$ characterizing particle cascades for $2 \leftrightarrow 2$ scattering associated to the phenomenon of weak wave turbulence. We show that this matches resummed perturbative estimates based on the two-particle irreducible (2PI) effective action (Sec.~\ref{sec:quantumanalytical}). In this perturbative regime we study the effect of quantum corrections and demonstrate that turbulence is absent at sufficiently short distances.

\section{Classical-statistical lattice gauge theory}
\label{sec:classicalsimulation}

Collisions of heavy nuclei leave behind a plasma of quarks and mostly
gluons in a nearly flat region of space because of Lorentz contraction along the beam- or $z$-axis. If the quarks and gluons streamed freely into the surrounding space, then the distribution of particles would become locally highly anisotropic. In this case the stress tensor quickly acquires an oblate form with $T_{xx} \sim T_{yy} \gg T_{zz}$. The fastest growing mode of plasma instabilities then has its wave vector along the normal direction and generates a prolate contribution to the stress, which pushes the system towards greater isotropy~\cite{Plasmainst,WongYangMills,Romatschke:2006nk,Berges:2007re}. 

We describe this physics using classical-statistical Yang-Mills theory in Minkowski space-time following closely Ref.~\cite{Berges:2007re}, to which we refer for further technical details.
It employs the Wilsonian lattice action for SU(N) gauge theory in Minkowski space-time:
\begin{eqnarray}\label{eq:LatticeAction}
S[U] &=& - \beta_0 \sum_{x} \sum_i \left\{ \frac{1}{2 {\rm Tr}
\mathbf{1}} \left( {\rm Tr}\, U_{x,0i} + {\rm Tr}\, U_{x,0i}^{\dagger}
\right) - 1 \right\}
\nonumber\\
&& + \beta_s \sum_{x} \sum_{i<j} \left\{ \frac{1}{2 {\rm
Tr} \mathbf{1}} \left( {\rm Tr}\, U_{x,ij} + {\rm Tr}\, U_{x,ij}^{\dagger}
\right) - 1 \right\} \, ,
\end{eqnarray}
with $x = (x^0, {\bf x})$ and spatial Lorentz indices $i,j = 1,2,3$. It is given in terms of the plaquette variable $U_{x,\mu\nu} \equiv U_{x,\mu} U_{x+\hat\mu,\nu}
U^{\dagger}_{x+\hat\nu,\mu} U^{\dagger}_{x,\nu}$,
where $U_{x,\nu\mu}^{\dagger}=U_{x,\mu\nu}\,$. Here $U_{x,\mu}$ is the
parallel transporter associated with the link from the neighboring
lattice point $x+\hat{\mu}$ to the point $x$ in the direction of
the lattice axis $\mu = 0,1,2,3$. The definitions
$\beta_0 \equiv 2 \gamma {\rm Tr} \mathbf{1}/g_0^2$ and
$\beta_s \equiv 2 {\rm Tr} \mathbf{1}/(g_s^2 \gamma)$
contain the lattice parameter $\gamma \equiv a_s/a_t$, where $a_s$ denotes the spatial and $a_t$ the temporal lattice spacings, and we will consider $g_0 = g_s = g$.              
             
We specify to SU(2) as the gauge group. The dynamics is solved in temporal axial gauge. Varying the action (\ref{eq:LatticeAction}) w.r.t.\ the spatial link variables $U_{x, j}$ yields the classical equations of motion. Variation w.r.t.\ to a temporal link gives the Gauss constraint. The coupling $g$ can be scaled out of the equations of motion and we will set $g = 1$ for the simulations. We define the gauge fields as\footnote{Here we use a different normalization of the gauge field than in equation (8) of Ref.~\cite{Berges:2007re}.}
\begin{equation}\label{eq:compute-gauge-field}
A_i^a(x) \,=\, -\frac{i}{a_s} \, {\rm Tr} \, \left( \sigma^a U_i(x) \right) 
\end{equation}
where $\sigma_a$ are the Pauli matrices with color index $a = 1, 2, 3$. Correlation functions are obtained by repeated numerical integration of the classical lattice equations of motion and Monte Carlo sampling of initial conditions. To be specific, we consider the extreme anisotropy case described by an effectively $\delta(p_z)$-like initial Gaussian distribution. The initial time derivatives $\dot{A}_{\mu}(t=0, \vec{x})$ are set to zero, which implements the Gauss constraint at all times. Shown results are from computations on $N^3 = 128^3$ lattices, where gauge fixing to Coulomb gauge is employed using a stochastic overrelaxation algorithm~\cite{Cucchieri:1995pn}. We comment on the observed stability of the results on $64^3$ and $256^3$ lattices below.

\begin{figure}[t]
\centerline{\includegraphics[scale=0.4,angle=-90]{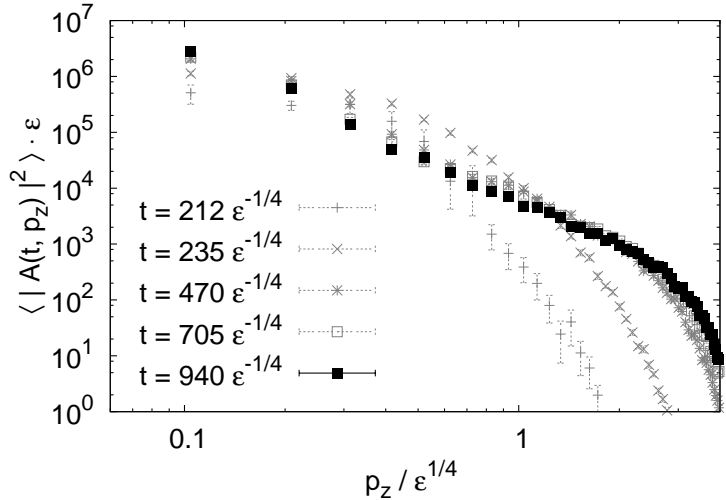}}
\caption{\label{fig:times} Fourier coefficients of the squared modulus of the gauge field in Coulomb gauge as a function of momentum for different snapshots in time.}
\end{figure}
Fig.~\ref{fig:times} shows the equal-time correlation function
\begin{equation}\label{eq:def-F-classical}
F^{{\rm (cl)} \, ab}_{\mu \nu}(t; {\bf p}) = \int d^3 x \, e^{- i {\bf p} \cdot {\bf x} } \langle \, A_{\mu}^a(t, {\bf x}) A_{\nu}^b(t, 0) \, \rangle 
\end{equation}
as a function of momentum for various instances of time in units of the energy density $\epsilon$.
The direction of the gauge fields is in the transverse $xy$-plane and the momenta are parallel to the $z-$axis. This correlation function is typically associated to an occupation number distribution divided by frequency. Because of the oblate initial conditions the distribution along the $z-$axis is characterized by small values at early times. Exponential growth due to plasma instabilities leads to a rapid population of long-wavelength modes, while fluctuation effects induced by the growth in the lower momentum modes lead to amplification in a broad range of momenta. The exponential growth stops around $t \simeq (200 \, - \, 250) \epsilon^{-1/4}$, and we concentrate on the subsequent evolution. The behavior at earlier times is analyzed in detail in Ref.~\cite{Berges:2007re}.  
\begin{figure}[t]
\centerline{\includegraphics[scale=0.4,angle=-90]{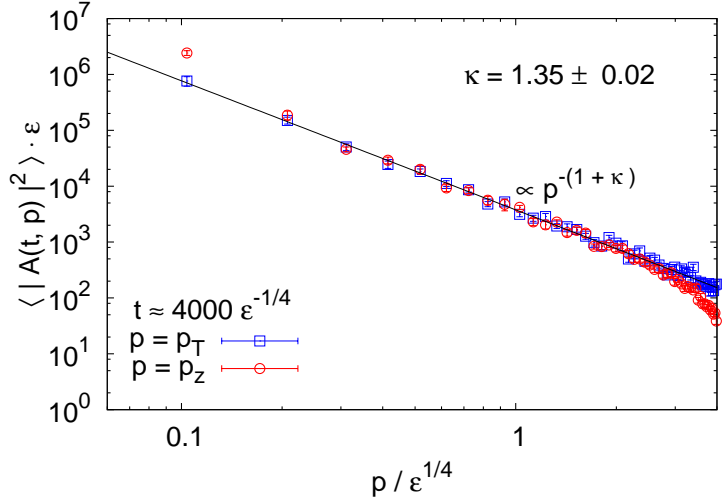}}
\caption{\label{fig:fits} Simulation results for transverse (square) and longitudinal momenta (circle) at a later time. The fit indicates a power-law behavior.}
\end{figure}

\begin{figure}[t]
\centerline{\includegraphics[scale=0.4,angle=-90]{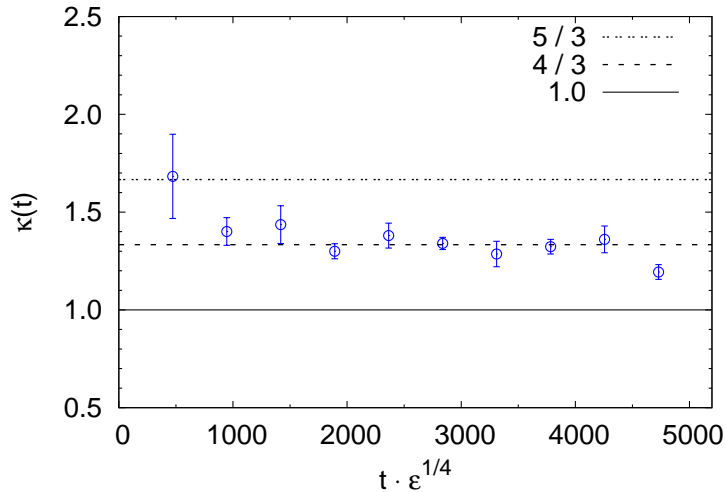}}
\caption{\label{fig:kappa-vs-t} Fits to a power-law behavior with exponent $\kappa$ taken at different times. The dashed lines represent perturbative Kolmogorov exponents.}
\end{figure}
After the exponential growth period the dynamics slows down considerably. As time proceeds, more and more ultraviolet modes approach a power-law described by
\begin{equation}
\langle |A(t,{\bf p})|^2 \rangle \sim |{\bf p}|^{-(1+\kappa)} \, ,
\label{eq:powerlaw}
\end{equation}
with the "occupation number exponent" $\kappa$. Fig.~\ref{fig:fits} shows the simulation results at time $t = 4000 \epsilon^{-1/4}$ for transverse (square) and longitudinal momenta (circle). Given that longitudinal and transverse momentum modes were populated very differently at initial time, the differences reflect the comparably small remaining anisotropy at this time. The solid line represents a power-law fit (\ref{eq:powerlaw}) with
\begin{equation}\label{eq:overall-kappa}
 \kappa = 1.35 \pm 0.02 \; ,
\end{equation}
which matches the data in a significant momentum range rather well. 

In order to obtain the fit value for $\kappa$ from the simulations, we first employ a least-square fit to the transverse and longitudinal results separately at a given time. Subsequently, we average the results for the transverse and longitudinal momenta. For the longitudinal momenta we choose the fit range $p_z \in [ 0.6 , 2 ]\, \epsilon^{1/4} $. For the transverse-momentum results, which display power-law behavior in a larger domain at earlier times, we use the range $ p_{\perp} \in [ 0.6 , 3 ]\, \epsilon^{1/4}$. Convergence in the infrared proceeds rather slowly. However, changing the lower bound for the fit range from $0.6 \epsilon^{1/4}$ down to $0.2 \epsilon^{1/4}$ leads only to changes in $\kappa$ which are comparable to the statistical error. 
Fig.~\ref{fig:kappa-vs-t} displays the fit values for $\kappa$ as a function of time. The comparably large error for $t \lesssim 1000 \epsilon^{-1/4}$ comes from the insufficient isotropization at these times. Subsequently, the values for $\kappa$ change very little as a function of time and the errors become small. The dashed lines correspond to the results for perturbative Kolmogorov exponents for energy ($5/3$) and particle ($4/3$) cascades, which is explained in Sec.~\ref{sec:quantumanalytical}. The late-time behavior for $t \gtrsim 4500 \epsilon^{-1/4}$ is consistent with a slow approach to a classical thermal equilibrium distribution with exponent one (solid line). A remarkable agreement between simulation results and $\kappa = 4/3$ is observed in the quasi-stationary period around $1000 \epsilon^{-1/4} \lesssim t \lesssim 4500 \epsilon^{-1/4}$. We obtain the best fit value (\ref{eq:overall-kappa}) by taking the time-average in this quasi-stationary period. 

We emphasize that the errors quoted are purely statistical.  The different dynamics for longitudinal and transverse modes leads to the fact that the longitudinal distribution corresponds to somewhat larger estimates for $\kappa$ than what is inferred for transverse modes at not too late times. If this difference is employed to estimate the systematic error for $\kappa$, the error will be less than ten per cent.

Our results seem to be incompatible with the value two for $\kappa$ reported in Refs.~\cite{Arnold:2005ef,Strickland:2007fm} based on Vlasov equations. Unfortunately, the origin of that value is not fully understood so far. In particular, a perturbative analysis in terms of Kolmogorov wave turbulence similar to the one in Sec.~\ref{sec:quantumanalytical} seems not to be able to explain it. There is still no consensus about the underlying reasons for the different numerical estimates of $\kappa$ using Vlasov and classical-statistical approaches, respectively. It has been pointed out that in the Vlasov treatment the hard modes represent static sources, which do not isotropize. In contrast, total energy is conserved in the classical-statistical lattice gauge theory simulation and all modes isotropize at sufficiently late times. However, for weak coupling and assuming a sufficiently large separation of scales both approaches are expected to agree. One could conclude that simulations on very much larger lattices to achieve the clear separation of scales underlying the Vlasov approach might be required to compare the two. We verified that simulations on lattices as big as $256^3$, with much less statistics, show no indication for quantitatively relevant changes of the $\kappa$ value. We found that already simulations on $64^3$ lattices accurately reproduce (\ref{eq:overall-kappa}), which indicates that the results are rather robust. 

Clearly, none of these approximations are sufficient to quantitatively address the asymptotic late-time behavior, which should finally be characterized by a Bose-Einstein distribution for the gluons. While for low momenta the employed classical-statistical simulations are expected to give an accurate description for sufficiently high energy densities or occupation numbers, the high-momentum behavior will be altered by quantum corrections. In the following, we address aspects of this question using resummed perturbation theory at high momenta.

\section{Resummed perturbative quantum theory}
\label{sec:quantumanalytical}

For quantum fields $A^a_{\mu}(x)$ one can define two independent two-point correlation functions out of equilibrium, which may be associated to the anti-commutator and the commutator
\begin{equation}
F_{\mu \nu}^{ab}(x,y) \,=\, \frac{1}{2} \langle \, \{ A^a_{\mu}(x), A^b_{\nu}(y) \} \, \rangle \quad , \quad \rho^{ab}_{\mu \nu}(x,y) \,=\, i \langle \, [ A^a_{\mu}(x), A^b_{\nu}(y) ] \, \rangle \, ,
\label{eq:def-F-quantum}
\end{equation}
respectively. Loosely speaking, the spectral function $\rho$ determines which states are available, while the statistical propagator $F$ contains the information about how often a state is occupied. The spectral function is related to the retarded propagator $G_{(R)}$ and the advanced one $G_{(A)}$ as $\rho = G_{(R)} - G_{(A)}$. A tremendous simplification of thermal equilibrium is that the spectral and statistical functions are related by the fluctuation-dissipation relation, which is not assumed here~\cite{Berges:2004yj}. 

In the following we will show how a power-law behavior (\ref{eq:powerlaw}) can be explained for sufficiently high momenta in resummed perturbation theory, if quantum corrections are neglected. Since the scaling solution is time and space translation invariant, the correlators in (\ref{eq:def-F-quantum}) can be Fourier transformed to $\tilde{F}(p)$ and $\tilde{\rho}(p)$ with four-momentum $p=(p^0,{\bf p})$.\footnote{We introduce a $-i$ in Fourier transforms of the spectral ($\rho$-) and retarded/advanced components, such as $\tilde{\rho}(p) = - i \int {\rm d}^4 x\, e^{ip_\mu x^\mu} \rho(x)$, while $\tilde{F}(p) = \int {\rm d}^4 x\, e^{ip_\mu x^\mu} F(x)$~\cite{Berges:2004yj}.} Similar to (\ref{eq:def-F-quantum}) we consider statistical, $\tilde{\Pi}_{(F)}$, and spectral, $\tilde{\Pi}_{(\rho)}$, components of self-energies defined as~\cite{Berges:2004yj}
\begin{eqnarray}
\tilde{\Pi}^{\mu \nu}_{{\rm (}F{\rm )} ab}(p) &=& \tilde{G}_{{(R)} ac}^{-1\, \mu \gamma}(p)\,
\tilde{F}_{\gamma \delta}^{cd}(p)\, \tilde{G}_{{(A)} db}^{-1\, \delta \nu}(p)  \, ,
\nonumber\\
\tilde{\Pi}^{\mu \nu}_{{\rm (}\rho{\rm )} ab}(p) &=& \tilde{G}_{{(R)} ab}^{-1\, \mu \nu}(p)
- \tilde{G}_{{(A)} ab}^{-1\, \mu \nu}(p) \, ,
\label{eq:self-energies}
\end{eqnarray}
where summation over repeated Lorentz and color indices is implied. The translation invariant propagators (\ref{eq:def-F-quantum}) and self-energies (\ref{eq:self-energies}) obey the nontrivial identity~\cite{Berges:2008sr}
\begin{equation}
\tilde{\Pi}^{\mu \nu}_{{\rm (}\rho{\rm )} ab}(p)\, \tilde{F}_{\nu \mu}^{ba}(p) - \tilde{\Pi}^{\mu \nu}_{{\rm (}F{\rm )} ab}(p)\, \tilde{\rho}_{\nu \mu}^{ba}(p) \,=\, 0 \, ,
\label{eq:identity}
\end{equation}
which can be directly verified by plugging in the above definitions. This equation is well-known in nonequilibrium physics and will be the starting point for our calculation. In the language of Boltzmann dynamics it states that "gain terms" equal "loss terms" for which stationarity is achieved~\cite{Berges:2004yj}. Thermal equilibrium trivially solves~\eqref{eq:identity}, which we do not consider in the following. Instead, we will look for possible non-thermal stationary solutions in perturbation theory relevant at sufficiently high momenta.

\begin{figure}[t]
\centerline{\includegraphics[scale=0.25,angle=0]{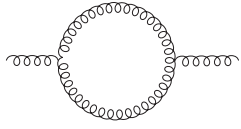}}
\caption{\label{fig:1-loop-diagram}  Gluon part of the one-loop contribution to the self-energy with (2PI) resummed propagator lines.}
\end{figure}
To exemplify the evaluation of the self-energies, we consider the one-loop diagram of Fig.~\ref{fig:1-loop-diagram} appearing in the two-particle irreducible (2PI) effective action scheme, where self-energies are expressed in terms of self-consistently dressed propagators while vertices remain undressed. Using $\tilde{F}^{ab}_{\mu\nu} = \delta^{ab} \tilde{F}_{\mu\nu}$ and $\tilde{\rho}^{ab}_{\mu\nu} = \delta^{ab} \tilde{\rho}_{\mu\nu}$ at one-loop order the gauge field contributions to the color-diagonal self-energies can be written as~\cite{Berges:2004pu}
\begin{eqnarray}
\tilde{\Pi}^{\mu \nu}_{{\rm (}F{\rm )}}(p) &=& - g^2 \int_{k q} (2\pi)^4\, \delta^{(4)}(p + k + q ) \left[ \tilde{F}_{\beta \alpha }(k) \tilde{F}_{\delta \gamma}(q) + \frac{1}{4} \tilde{\rho}_{\beta \alpha}(k) \tilde{\rho}_{\delta \gamma}(q) \right] 
\nonumber\\
&& \times \, V^{\mu \alpha \gamma , \nu \beta \delta}(p,k,q) \, ,
\nonumber\\
\tilde{\Pi}^{\mu \nu}_{{\rm (}\rho{\rm )}}(p) & = & g^2 \int_{k q} (2\pi)^4\, \delta^{(4)}(p + k + q ) \left[ \tilde{F}_{\beta \alpha}(k) \tilde{\rho}_{\delta \gamma}(q) + \tilde{\rho}_{\beta \alpha}(k) \tilde{F}_{\delta \gamma}(q) \right] 
\nonumber\\
&& \times \, V^{\mu \alpha \gamma , \nu \beta \delta}(p,k,q) \, ,
\label{eq:one-loop}
\end{eqnarray}
with the notation $\int_k \equiv \int d^4 k/(2\pi)^4$.
The Lorentz and momentum structure of the two three-vertices appearing in the one-loop expression for SU(2) gauge theory is contained in
\begin{eqnarray}
V^{\mu \alpha \gamma , \nu \beta \delta}(p,k,q) &=& \left[ \, g^{\mu \alpha}(p - k)^{\gamma} +  g^{\alpha \gamma} (k -q)^{\mu} + g^{\mu \gamma } (q - p)^{\alpha} \, \right] 
\nonumber\\
& \times & \left[ \, g^{\nu \beta}(p-k)^{\delta} + g^{\beta \delta}(k-q)^{\nu} + g^{\nu \delta } (q-p)^{\beta}  \, \right] \, .
\end{eqnarray}

In accordance with (\ref{eq:powerlaw}), we are looking for scaling solutions which behave as
\begin{equation}
\tilde{F}_{\mu \nu}(s p ) = |s|^{- (2 + \kappa )} \tilde{F}_{\mu \nu}(p) \quad , \quad \tilde{\rho}_{\mu \nu}(s p) = |s|^{-2}\,{\rm sgn(s)}\, \tilde{\rho}_{\mu \nu}\left(p \right)  \, 
\label{eq:scaling-assumption}
\end{equation}
under rescaling with the real parameter $s$. The spectral function is antisymmetric, $\tilde{\rho}_{\mu \nu}(p) = - \tilde{\rho}_{\nu \mu}(-p)$, while $\tilde{F}_{\mu \nu}(p) = \tilde{F}_{\nu \mu}(-p)$ and we assume symmetry in Lorentz indices. As in Ref.~\cite{Berges:2008wm}, scaling solutions can be efficiently identified by integrating (\ref{eq:identity}) over external spatial momentum ${\bf p}$ and suitable scaling transformations for coordinates. This is similar to what is done in the context of weak Kolmogorov wave turbulence for scalar models using a Boltzmann equation~\cite{Zakharov}, and in this way the problem is reduced to simple algebraic conditions for exponents. Here, the analysis is complicated by the fact that for the perturbative result to order $g^2$ the two terms in (\ref{eq:identity}) vanish separately on-shell, i.e.\ for $p^0 = \pm |{\bf p}|$. The non-vanishing contributions of the diagram in Fig.~\ref{fig:1-loop-diagram} to order $g^4$ can be conveniently analyzed by writing (\ref{eq:self-energies}) as 
\begin{equation}\label{eq:F-GPiG-Fourier}
\tilde{F}_{\mu \nu}(p) = \tilde{G}^{(R)}_{\mu \alpha}(p) \tilde{\Pi}_{(F)}^{\alpha \beta} (p) \tilde{G}^{(A)}_{\beta \nu}(p) \quad , \quad \tilde{\rho}_{\mu \nu}(p) = \tilde{G}^{(R)}_{\mu \alpha}(p) \tilde{\Pi}_{(\rho)}^{\alpha \beta} (p) \tilde{G}^{(A)}_{\beta \nu}(p) \, .
\end{equation}
This is used to replace one propagator line in the one-loop graph of Fig.~\ref{fig:1-loop-diagram}. To order $g^4$, the self-energies appearing in (\ref{eq:F-GPiG-Fourier}) can then be taken as (\ref{eq:one-loop}). Following a well-trodden path~\cite{Berges:2008wm,Berges:2008sr,Micha:2004bv,Zakharov}, we obtain by integrating over spatial momentum ${\bf p}$ from the stationarity condition (\ref{eq:identity}):
\begin{eqnarray}
 0 & = & g^4 \int_{{\bf p}qklr} \!\!\!\! \delta^{(4)}(p + k + q)\delta^{(4)}(k + l + r) 
V_{\mu \delta \gamma , \nu \tau \lambda}(p,k,q) V_{\alpha \delta' \gamma' , \beta \tau' \lambda'}(k,l,r)
\nonumber\\
& \times & \tilde{G}_{(R)}^{\tau \alpha}(k)\tilde{G}_{(A)}^{\beta \delta}(k) \tilde{\rho}^{\lambda \gamma}(q) \tilde{F}^{\nu \mu}(p) \left\{ \tilde{F}^{\tau' \delta'}(l) \tilde{F}^{\lambda' \gamma'}(r) \left[ \left| \frac{p^0}{r^0} \right|^{\Delta} {\rm sgn}\left( \frac{p^0}{r^0} \right) \right. \right.
\nonumber\\
& + &   \left. \left| \frac{p^0}{l^0} \right|^{\Delta} {\rm sgn}\left( \frac{p^0}{l^0} \right) - \left| \frac{p^0}{q^0} \right|^{\Delta} {\rm sgn}\left( \frac{p^0}{q^0} \right) - 1 \right] + \frac{1}{4}\,
\tilde{\rho}^{\tau' \delta'}(l) \tilde{\rho}^{\lambda' \gamma'}(r)
\nonumber\\
& \times & \left. \left[ \left| \frac{p^0}{r^0} \right|^{\tilde{\Delta}} {\rm sgn}\left( \frac{p^0}{r^0} \right)
+ \left| \frac{p^0}{l^0} \right|^{\tilde{\Delta}} {\rm sgn}\left( \frac{p^0}{l^0} \right) - \left| \frac{p^0}{q^0} \right|^{\tilde{\Delta}} {\rm sgn}\left( \frac{p^0}{q^0} \right) - 1 \right] \right\}
. \qquad 
\label{eq:zeros}
\end{eqnarray}
Here 
\begin{equation}
\Delta = 4 - 3 \kappa
\label{eq:delta}
\end{equation}
with $\tilde{\Delta} \equiv \Delta + 2 \kappa$. In (\ref{eq:zeros}) the first term in brackets, $\sim \tilde{F}\tilde{F}$, vanishes for $\Delta = -1$, and for $\Delta = 0$ in the on-shell limit for $\tilde{F}$ and $\tilde{\rho}$ such that only $2 \leftrightarrow 2$ scattering contributes. With (\ref{eq:delta}) one observes that these correspond to the well-known Kolmogorov exponents for energy and for particle cascades,
\begin{equation}
\kappa = \frac{5}{3} \quad ,{\rm or} \quad \kappa = \frac{4}{3} \, ,
\label{eq:kappa_cl}
\end{equation}
respectively. In (\ref{eq:zeros}) the second term in brackets, $\sim \tilde{\rho}\tilde{\rho}$, does not vanish with (\ref{eq:kappa_cl}). Doing the equivalent calculation starting from the classical-statistical gauge theory instead of the quantum theory leads to the same stationarity equation (\ref{eq:zeros}), however, without this term $\sim \tilde{\rho}\tilde{\rho}$~\cite{Berges:2004pu}. Correspondingly, neglecting quantum corrections one observes rather good agreement of the perturbative solution $\kappa = 4/3$ with the value (\ref{eq:overall-kappa}) indicated by the full numerical solution of the classical-statistical gauge theory. A similar analysis can be performed for the two-loop gluon diagrams. 

For sufficiently high momenta, where characteristic occupation numbers are of order one, the quantum corrections appearing in (\ref{eq:zeros}) become relevant, which preempts a scaling solution (\ref{eq:kappa_cl}) in the far UV. On the other hand, parametrically the perturbative estimate certainly breaks down for occupancies as large as ${\cal O}(1/g^2)$, where order-one corrections to the self-energies occur at any order in a loop expansion. As a consequence, one expects a window of momenta for which the scaling solution (\ref{eq:kappa_cl}) can describe quantitatively the dynamics. 
In view of this, it is remarkable that the simulation results shown in Fig.~\ref{fig:fits} exhibit for low momenta only small deviations from the perturbative behavior. Given that the numerical results are rather insensitive to going to $256^3$ lattices, simulations on much larger lattices might be required to settle the important question of whether a new nonperturbative infrared fixed point exists in nonequilibrium QCD, similar to what is observed in scalar theories~\cite{Berges:2008wm}.

\section{Conclusions}

Despite important differences the discussion of the thermalization process in heavy-ion collisions and cosmology after inflation shows remarkable similarities. In both cases nonequilibrium instabilities can lead to a fast period of exponential growth of occupation numbers. This is followed by a slow period, where the quantitative agreement concerns characteristic exponents or scaling functions. Our numerical as well as analytical results indicate that the nonabelian gauge theory can lead to perturbative Kolmogorov scaling exponents known from scalar models. 
In QCD simulations for large volumes an ambitious analysis in terms of gauge invariant quantities such as stress-tensor correlation functions would be desirable for a quantitative understanding of the nonperturbative infrared dynamics. One can hope that this leads to an improved theoretical understanding of the plasma's temporal evolution, which is crucial for the successful outcome of the experimental heavy-ion program.\\

This work is supported in part by the BMBF grant 06DA267, and by the
DFG under contract SFB634. Some of the ideas were initiated during the program 
on "Nonequilibrium Dynamics in Particle Physics and
Cosmology" (2008) at the Kavli Institute for Theoretical Physics in Santa Barbara,
supported by the NSF under grant PHY05-51164.

\end{document}